\documentclass[fleqn,usenatbib]{mnras}
\usepackage{newtxtext,newtxmath}
\usepackage[T1]{fontenc}

\usepackage{graphicx}	
\usepackage{amsmath}	
\usepackage{pdflscape}	
\usepackage{bm} 

\title[orbital structure of planetary systems]{Orbital structure of planetary systems formed by giant impacts: stellar mass dependence}

\author[Hoshino \& Kokubo]{
H. Hoshino,$^{1,2}$
E. Kokubo,$^{2,1}$
\\
$^{1}$Department of Astronomy, The University of Tokyo, Hongo, Bunkyo-ku, Tokyo 113-0033, Japan\\
$^{2}$Division of Science, National Astronomical Observatory of Japan, Osawa, Mitaka, Tokyo 181-8588, Japan
}

\date{Accepted 2022 December 16. Received 2022 December 13; in original form 2022 September 30}

\pubyear{2022}

\begin{document}
\label{firstpage}
\pagerange{\pageref{firstpage}--\pageref{lastpage}}
\maketitle

\begin{abstract}
Recent exoplanet surveys revealed that for solar-type stars, close-in Super-Earths are ubiquitous and many of them are in multi-planet systems.
These systems are more compact than the Solar System's terrestrial planets.   
However, there have been few theoretical studies on the formation of such planets around low-mass stars.
In the standard model, the final stage of terrestrial planet formation is the giant impact stage, where protoplanets gravitationally scatter and collide with each other and then evolve into a stable planetary system. 
We investigate the effect of the stellar mass on the architecture of planetary systems formed by giant impacts.
We perform {\it N}-body simulations around stars with masses of 0.1--2 times the solar mass.
Using the isolation mass of protoplanets, we distribute the initial protoplanets in 0.05--0.15 au from the central star and follow the evolution for 200 million orbital periods of the innermost protoplanet.
We find that for a given protoplanet system, the mass of planets increases as the stellar mass decreases, while the number of planets decreases.
The eccentricity and inclination of orbits and the orbital separation of adjacent planets increase with decreasing the stellar mass.
This is because as the stellar mass decreases, the relative strength of planetary scattering becomes more effective.
We also discuss the properties of planets formed in the habitable zone using the minimum-mass extrasolar nebula model.
\end{abstract}




\section{Introduction} \label{sec:intro}

Since 1995, over 5000 exoplanets have been found \citep[e.g.,][]{Winn_2015, Zhu_2021} .
Many of them rotate around solar-type stars and it is currently recognized that the most common type of  planet is Super-Earths.
Super-Earths are several times more massive than the Earth.
They are found in more compact and close-in systems than the solar system terrestrial planets.
We need a more general theory of planet formation that can explain these systems.

Several projects searching for Earth-like planets focusing on M-type stars have been undertaken.
An M-type star is the smallest and coolest kind of stars on the main sequence, with masses of about 0.1--0.5 solar masses.
The Subaru InfraRed Doppler \citep[IRD;][]{Tamura_2012} project
is looking for habitable planets by observing the infrared light, which is emitted more strongly than visible light by M-stars.
Other projects include CARMENES \citep{Quirrenbach_2010}, HPF \citep{Mahadevan_2010}, and SPIRou \citep{Micheau_2012}. These projects have already observed exoplanets around low-mass stars \citep[e.g.,][]{Kaminski_2018,Harakawa_2022}.
Although the observational data of planets around M-type stars are currently insufficient, M-type stars are known to be the most common stars \citep[e.g.,][]{Bochanski_2010} and planet occurrence rates tend to be high around low mass stars \citep[e.g.,][]{Hardegree-Ullman_2019, Dressing_2013, Dressing_2015, Yang_2020, He_2021},
and hence we expect that more planets will be found around M-type stars.
It is therefore important to study the formation of terrestrial planets around low-mass stars.

Planets are formed by dust growth in the protoplanetary disk.
The dust first aggregates to form planetesimals.
Protoplanets are then formed through runaway and oligarchic growth of planetesimals \citep[][]{Kokubo_1998,Kokubo_2000,Kokubo_2002}.
Another model of this growth process is the accretion of cm-sized pebbles \citep[e.g.,][]{Ormel-Klahr_2010, Lambrechts-Johansen_2012}.
Finally, protoplanets collide and merge through orbital crossing, and form terrestrial planets \citep[e.g.,][]{Hayashi_1985, Kokubo_2012, Raymond_2014}.
This final process is known as the giant impact stage \citep[e.g.,][]{Hartmann_1975, Wetherill_1990, Kokubo_2006}.
Previous studies investigated the influence of the initial protoplanet systems such as the total disk mass, disk radial profile, orbital separation, and distance from the central star \citep[e.g.,][]{Wetherill_1996, Raymond_2005, Kokubo_2006, Raymond_2007}.
However, a stellar mass of 1 solar mass was used in most cases.

Clarifying the influence of the stellar mass on the planetary system architecture is important for understanding the diversity of planetary systems.
There have been few theoretical studies focusing on the stellar mass.
Using {\em N}-body simulations, \citet{Raymond_2007} and \citet{Ciesla_2015} studied the masses and water mass fractions of planets forming in the habitable zone (HZ) by varying the stellar mass.
The HZ is the area where liquid water can exist on the planet surface and is represented as a distance from the central star.
These two studies adopted stellar masses of $M_*=0.2$--$1M_{\odot}$. 
\citet{Moriarty_2016} considered $0.2 M_\odot$ and $1\, M_{\odot}$ stars and compared their results with Kepler planets.
These studies revealed that there is a positive correlation between the planet mass and host star's mass,
consistent with some observational results \citep[e.g.,][]{Wu_2019, Pascucci_2018}.
In addition, \citet{Ansdell_2017} reported that the dust mass in protoplanetary disks has a positive correlation with the stellar mass.
\cite{Mulders_2021} discussed how the mass distribution of close-in super-Earths changes with different stellar masses considering the presence of outer giant planets and the effects of pebble accretion.
It is clear that the mass of a central star affects planetary system formation and evolution.
However, no previous studies examined the stellar mass dependence of planetary system architecture systematically. 

Given this trend, it is important to investigate the formation of short-period planets around low-mass stars.
The dynamical properties of planetary systems formed in the vicinity of low-mass stars can be different from those around 1 au of the solar mass star. 
The gravitational interaction among planets can be scaled using the Hill radius of planets that depends not only the planet mass but also the stellar mass and the orbital radius \citep{Nakazawa_1988}.
However, if collisions among planets are included, the Hill scaling should break down since another length scale, the physical radius of planets, that is not scaled by the Hill radius comes in.
Therefore, it is necessary to confirm what actually happens in the vicinity of low-mass stars in the giant impact stage.
As we have described so far, it is expected that terrestrial planets including Earth-mass planets are found around low-mass stars.
In this study, we would like to understand how the architecture of planetary systems changes depending on the stellar mass, which will lead to understanding the formation of terrestrial planets in general.

We investigate the giant impact stage of planetary system formation by systematically changing the stellar mass and its effect on the final orbital architecture of planetary systems using $\it{N}$-body simulations.
We vary the stellar mass from 0.1 to 2 solar mass.
We explain our calculation models in Section 2 and then show our results in Section 3.
Section 4 is devoted to a summary and discussion.

\section{Models and Methods}

\begin{table}
 \setlength{\tabcolsep}{3pt}
 \caption{Initial conditions of protoplanet systems}
 \label{tab:condition}
 \begin{tabular}{cccccccccc}
  \hline
  Model & $M_{*}$ & $\Sigma_0$ & $\alpha$ & $\beta$ & $a_{\rm{in}}$ & $a_{\rm{out}}$ & $\tilde{b}_{\rm{ini}}$ & $\it{N}_{\rm{ini}}$ & $m_{\rm{tot}}$\\
   & [$M_{\sun}$] & $[\rm{g\, cm^{-2}}]$ & & & [au] & [au] & & & $[m_{\earth}]$\\
  \hline
    S1 & 2 & 10 & -1.5 & 0 & 0.05 & 0.15 & 10 & 42 & 0.758 \\
    S2 & 1 & 10 & -1.5 & 0 & 0.05 & 0.15 & 10 & 30 & 0.767 \\
    S3 & 0.5 & 10 & -1.5 & 0 & 0.05 & 0.15 & 10 & 21 & 0.751 \\
    S4 & 0.2 & 10 & -1.5 & 0 & 0.05 & 0.15 & 10 & 14 & 0.804 \\
    S5 & 0.1 & 10 & -1.5 & 0 & 0.05 & 0.15 & 10 & 10 & 0.806 \\
    S6 & 0.2 & 20 & -1.5 & 0 & 0.05 & 0.15 & 10 & 10 & 1.61 \\
    S7 & 0.2 & 50 & -1.5 & 0 & 0.05 & 0.15 & 10 & 6 & 3.63 \\
    S8 & 0.2 & 10 & -1.5 & 0 & 0.05 & 0.15 & 7.5 & 21 & 0.781 \\
    S9 & 0.2 & 10 & -1.5 & 0 & 0.05 & 0.15 & 5 & 38 & 0.770 \\
    \hline
    E1 & 2 & 50 & -1.76 & 1.39 & 0.05 & 0.15 & 10 & 9 & 19.9 \\
    E2 & 1 & 50 & -1.76 & 1.39 & 0.05 & 0.15 & 10 & 10 & 7.34 \\
    E3 & 0.5 & 50 & -1.76 & 1.39 & 0.05 & 0.15 & 10 & 12 & 2.98 \\
    E4 & 0.2 & 50 & -1.76 & 1.39 & 0.05 & 0.15 & 10 & 14 & 0.810 \\
    E5 & 0.1 & 50 & -1.76 & 1.39 & 0.05 & 0.15 & 10 & 16 & 0.309 \\
    \hline
    EH1 & 1 & 50 & -1.76 & 1.39 & 0.625 & 1.68 & 10 & 7 & 12.5 \\
    EH2 & 0.5 & 50 & -1.76 & 1.39 & 0.110 & 0.294 & 10 & 10 & 3.22 \\
    EH3 & 0.2 & 50 & -1.76 & 1.39 & 0.0403 & 0.108 & 10 & 13 & 0.681 \\
    EH4 & 0.1 & 50 & -1.76 & 1.39 & 0.0187 & 0.0502 & 10 & 16 & 0.212 \\
  \hline
 \end{tabular}
\end{table}

To investigate the dependence of the architecture of planetary systems on the stellar mass, we conduct $\it{N}$-body simulations of the giant impact stage around stars with various masses. 
First we describe the assumptions of our models and explain the initial conditions.
Then we show the method of the $\it{N}$-body simulations. 
Finally, we explain how to evaluate the results.

\subsection{Initial Conditions}

\subsubsection{Disk Models}

Since the giant impacts of protoplanets take place after gas dispersal, we consider gas-free disks.
We assume a solid surface density of protoplanetary disks of
\begin{equation}
 \Sigma=\Sigma_0 \left(\frac{a}{1\, \rm{au}}\right)^{\alpha} \left(\frac{M_*}{M_{\odot}}\right)^{\beta} \, \, \rm{g\, cm^{-2}},\label{equ:surface_density} 
\end{equation}
where $\Sigma _0$ is the reference surface density at 1 au around a 1-solar-mass star and  $\alpha$ and $\beta$ are the power-law indexes of the density profile and stellar mass dependence, respectively. 
We adopt a power-law disk model similar to the minimum-mass solar nebula \citep[MMSN][]{Hayashi_1981} and the minimum-mass extrasolar nebula (MMEN) models to produce initial protoplanet distributions.
With this model we can use the disk parameters for the global properties of the protoplanet distributions.
Since we focus on the effect of stellar mass on planetary systems, we use the most basic disk model. 
We set $\alpha =-3/2\,$ in accordance with the MMSN.
A disk with $\Sigma _0 \simeq 50$ and $\beta = 0$ corresponds to MMEN constructed by \cite{Chiang_2013} based on exoplanets discovered by the Kepler mission. 
\cite{Dai_2020} extended the MMEN model to include the stellar mass with $\beta \simeq 1$.
We consider $\Sigma _0 =10, \,20, \,50\, \rm{g}\,\rm{cm}^{-2}$, $\alpha =-1.5, \,-1.76$, and $\beta=0,\, 1.39$.
In the following, we call the disk with $\Sigma _0$ = 10 $\rm{g}\,\rm{cm}^{-2}$, $\alpha =-1.5$, and $\beta = 0$ the standard disk model and the disk with $\Sigma _0$ = 50 $\rm{g}\,\rm{cm}^{-2}$, $\alpha =-1.76$,  and $\beta$ = 1.39 the MMEN model. We adopt Equation (22) of \cite{Dai_2020} as the parameters of the MMEN model.

\subsubsection{Protoplanets} 

We assume that protoplanets are formed from a planetesimal disk with a surface density given by Eq.\ (\ref{equ:surface_density}) with certain orbital intervals. 
We adopt the oligarchic growth model \citep[e.g.,][]{Kokubo_1998, Kokubo_2002} that assumes that the orbital separation of adjacent protoplanets $b$ is proportional to the mutual Hill radius,
\begin{equation}
r_{\rm{H},\, \it{j}} = \left( \frac{m_j + m_{j+1}}{3 M_*} \right) ^{1/3} \frac{a_j + a_{j+1}}{2},
\label{equ:Hill_radius}
\end{equation}
where $m$ and $a$ are the mass and semimajor axis of the protoplanets.
Based on the oligarchic growth model, the mass of protoplanets is estimated to be
\begin{eqnarray}
m_{\rm{iso}} &\simeq& 2	\pi a b \Sigma \nonumber\\
&=&0.16 \left( \frac{\tilde{b}}{10} \right) ^{3/2} \left(\frac{\Sigma_0}{10}\right)^{3/2} \left(\frac{a}{1 \, \rm{au}}\right)^{(3/2)(2+\alpha)} \nonumber\\
&\,\,&\times \left(\frac{M_*}{M_{\odot}}\right) ^{(3/2)(\beta-1/3)} m_{\oplus},
\label{m_iso}
\end{eqnarray}
where $\tilde{b}$ is the initial orbital separation scaled by the mutual Hill radius $\tilde{b}=b/r_{\rm{H},\, \it{j}}$ and $m_{\oplus}$ is the Earth mass.
Although \cite{Kokubo_2006} found there is no dependence of the mass distribution of planets on the initial orbital separation, we vary $\tilde{b}=5, \,7.5, \,10$ to confirm the dependence. 
Planetary systems around low-mass stars are more likely to exist closer to the star than those around 1 $M_{\odot}$ stars\citep[e.g.,][]{Raymond_2007, Ciesla_2015}.
Therefore, we focus on the region 0.05--0.15 au. 
The eccentricity $e$ and inclination $i$ of protoplanets are distributed by the Rayleigh distribution.
We set the dispersions so that $\left< \tilde{e}^2 \right> ^{1/2} =  2 \left< \tilde{i}^2 \right> ^{1/2}= 1$, where $\tilde{e} $ and $\tilde{i}$ are the eccentricity and inclination scaled by the reduced Hill radius given by

\begin{equation}
h = \left( \frac{m}{3 M_*} \right) ^{1/3}.
\label{equ:reduced_Hill_radius}
\end{equation}
This value is small enough because the initial orbital spacing is 10 $r_{\rm{H},\, \it{j}}$. 
We confirmed that the results remain the same when the initial $e$ and $i$ are doubled.
The other angular orbital elements are given randomly. 
We perform 20 runs per model with different initial angular distributions of protoplanets. 
 
\subsubsection{Stellar Mass and Habitable Zone}

We also investigate the planet formation in HZ.
We consider stellar masses in the range $M_*$ = $0.1$--$1 \, M_{\odot}$. 
The HZ depends on the stellar mass. 
For the Solar System, HZ is estimated to be 0.8--1.5 au. 
For other stars, we calculate HZ using the mass $M_*$-luminosity $L_*$ relation given in \citet{Scalo_2007}: 
 \begin{eqnarray}
\log \frac{L_*}{L_{\odot} }=4.101\left( \log  \frac{M_*}{M_{\odot}} \right)^3 &+& 8.162\left(\log\frac{M_*}{M_{\odot}}\right)^2\nonumber\\
+ 7.108\left(\log\frac{M_*}{M_{\odot}}\right)&+&0.065,
\end{eqnarray} 
 where $L_{\odot}$ is the solar luminosity. 
This fitting equation can be used only in the range $M_*=$ 0.1--1.0 $M_{\odot}$.
 After calculating the luminosity, we obtain the HZ using the scaling law,
  \begin{equation}
\left( \frac{L_*}{L_{\odot}} \right) ^{1/2}=\frac{a_*}{a_{\odot}}, 
\label{HZ_eq}
\end{equation} 
where $a_*$ and $a_{\odot}$ are distances from the star and the Sun, respectively. 
In our model, we distribute protoplanets over a range of 1.5 times the HZ width and follow their evolution for 200 million periods at the HZ inner edge.
We only use planets whose final positions are within the HZ in our analysis.
If there is only one planet in the HZ and the orbital separation of the planets cannot be defined, then the width of the HZ is used as the separation instead.
We summarize the initial conditions of all models in Table \ref{tab:condition}.
Note that as the stellar mass changes, the linked parameters of protoplanets such as the mass and the Hill radius change.

\subsection{Orbital Integration}

The equation of motion of protoplanet $j$ is
\begin{equation}
\frac{\mathrm{d} \bm{v}_j}{\mathrm{d}t}=-GM_* \frac{\bm{x}_j}{ {\vert \bm{x}_j \vert}^3 } - \sum_{k \neq j}^N Gm_k \frac{\bm{x}_j-\bm{x}_k}{{\vert \bm{x}_j-\bm{x}_k \vert}^3},
\end{equation}
where $G$ is the gravitational constant and $\bm{v}_j$ and $\bm{x}_j$ are the velocity and position of protoplanet $j$, respectively. 
The first term is the gravity of the central star and the second term is the mutual gravitational interaction between protoplanets. 
We calculate the gravity directly by summing the interactions of all pairs and integrating the orbits of protoplanets with the fourth-order Hermite scheme \citep{Makino_1992, Kokubo_2004} with block timesteps \citep{Makino_1991}.
In addition, we implement the $\rm{P(EC)^{2'}}$ scheme \citep{Kokubo_1998_MNRAS} to reduce the integration error.

When two bodies' radii overlap, we assume perfect accretion in which they always merge. 
As shown in \citet{Kokubo_2010}, the assumption of perfect accretion barely affects the final structure of planetary systems.
Furthermore, \cite{Wallace_2017} and \cite{Esteves_2022} also showed that perfect accretion is sufficient for close-in planetary systems.
After a collision, a new particle is created with a mass equal to the sum of the two colliding particles and the position and velocity of the center of mass. 
The physical radius of a particle $r_{\rm{P}}$ is calculated by
\begin{equation}
r_{\rm{p}}= \left(\frac{3 m}{4\pi \rho}\right)^{1/3} ,
\end{equation}
where $\rho$ is the protoplanet's bulk density. 
We set $\rho = 3\, \rm{g} \, \rm{cm}^{-3}$ and fix it during the simulation.
The simulations are followed for 200 million $t_\mathrm{K}$, where $t_\mathrm{K}$ is 1 orbital period of the innermost protoplanet.
In this time-scale, the giant impact stage finishes and only a few final planets remain. 

\subsection{Orbital Parameters of Planetary Systems}

For a planetary system formed by giant impacts, we calculate the system parameters that reflect its orbital architecture: mass-weighted eccentricity $e_{\rm{m}}$, mass-weighted inclination $i_{\rm{m}}$, mean orbital separation $b$, and normalized angular momentum deficit (AMD) $D$ \citep[e.g.,][]{Laskar_1997}, given by 
\begin{eqnarray}
   e_{\rm{m}} & = & \frac{\Sigma_j^n\, e_j m_j}{\Sigma_j^n\, m_j},\\
   i_{\rm{m}} & = & \frac{\Sigma_j^n\, i_j m_j}{\Sigma_j^n\, m_j}, \\
   b & = & \frac{\Sigma_j^{n-1} (a_{j+1} - a_j)}{n-1}, \\
   D &=& \frac{\Sigma_j^n\, m_j \sqrt{a_j}\left( \rm{1}-\it{\cos{i_j}} \sqrt{\rm{1}-\it{e^{\rm{2}}_j}}\,\right)}{\Sigma_j^n\, m_j \sqrt{a_j}} .\label{equ:AMD}
\end{eqnarray}
The AMD $D$ represents the difference in orbital angular momentum from the planar and circular orbits.
Also, the mass and the number of final planets are calculated. 
Using the results of 20 runs per model, we investigate the relationship between the average of these values and the stellar mass. 

\section{Results}

\begin{figure*}
\centering
\includegraphics[width=17cm]{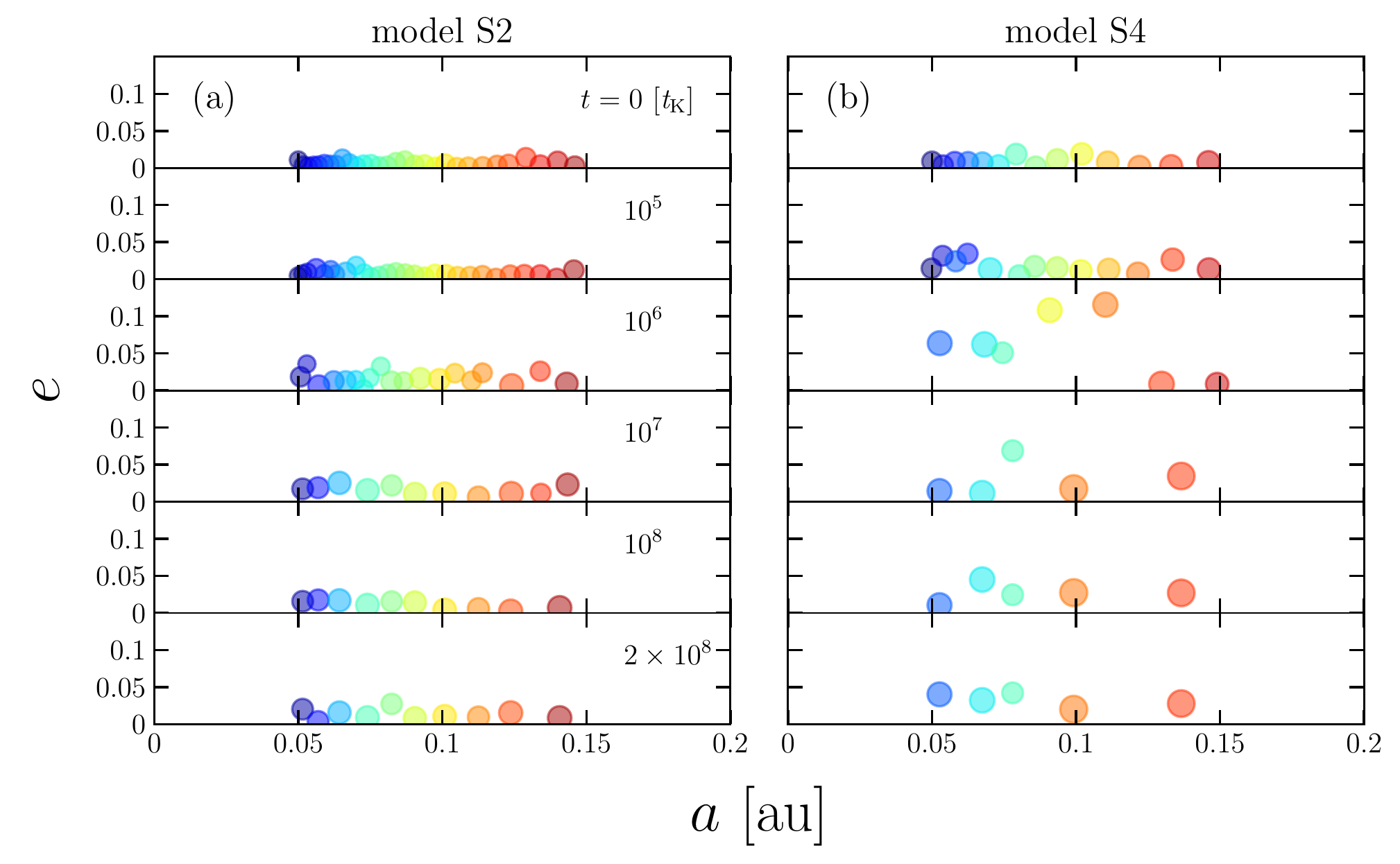}
\caption{Snapshots of the system on the eccentricity--semi-major axis plane at $t=0$, $10^5$, $10^6$, $10^7$, $10^8$, $2\times10^8$ $t_{\rm{K}}$. The left panel (a) is the case $M_* = 1 M_{\rm{\odot}}$ and the right panel (b) is that of  $M_* = 0.2 M_{\rm{\odot}}$ as a typical mass of an M-type star. $t_{\rm{K}}$ is one orbital period of the planet which has the closest orbit, 0.05 au at initial conditions. The size of each circle is proportional to the physical radius of the planet. The color corresponds to each particle.}
\label{fig:e-a_snap}
\end{figure*}

\begin{figure*}
\centering
\includegraphics[width=17cm]{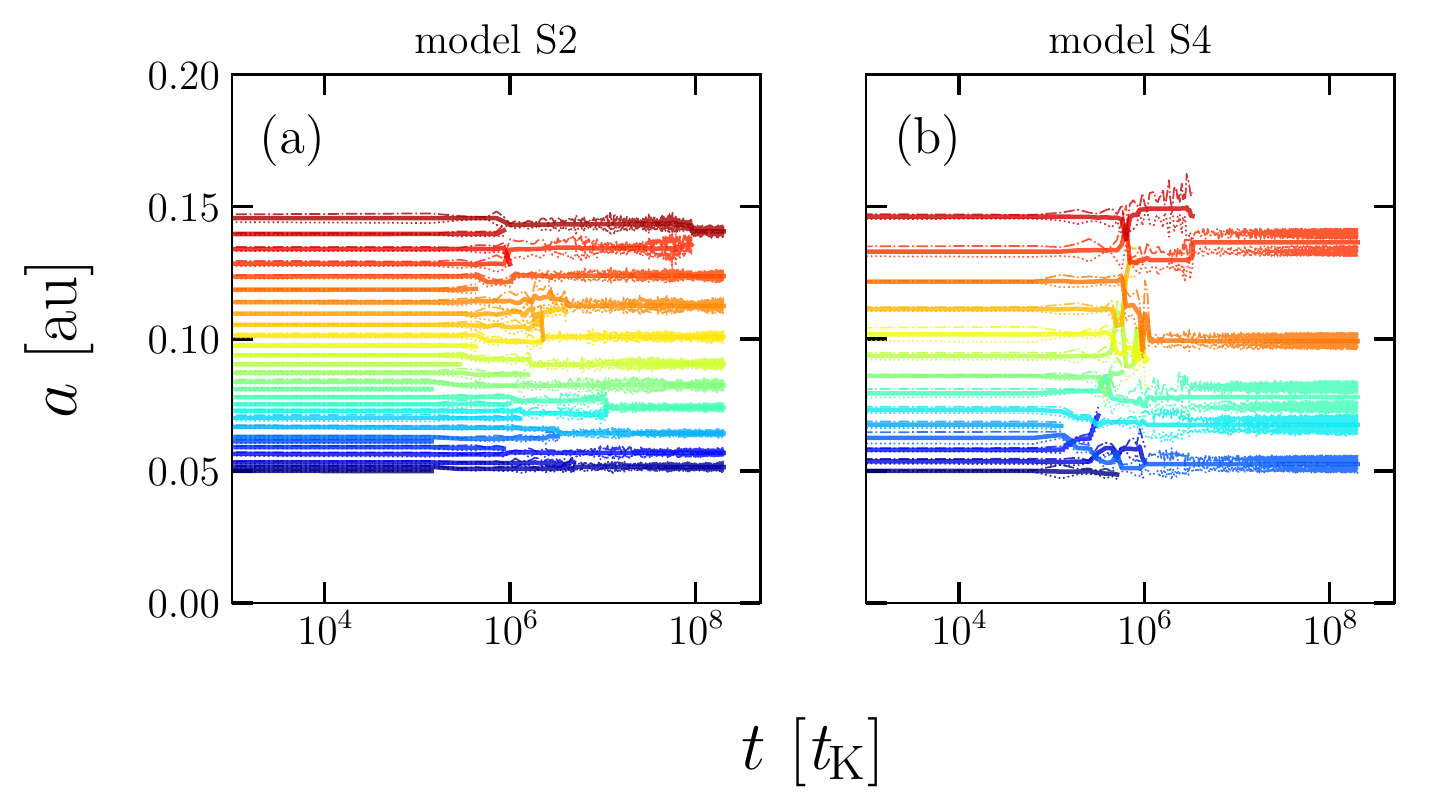}
\caption{Time evolution of the semi-major axis (solid line) and distances of periapsis (dotted line) and apoapsis (dash-dotted line) for the same run as in Fig. \ref{fig:e-a_snap} for model S2 (a) and model S4 (b).}
\label{fig:a-t_2-4}
\end{figure*}

\begin{figure*}
\centering
\includegraphics[width=16cm]{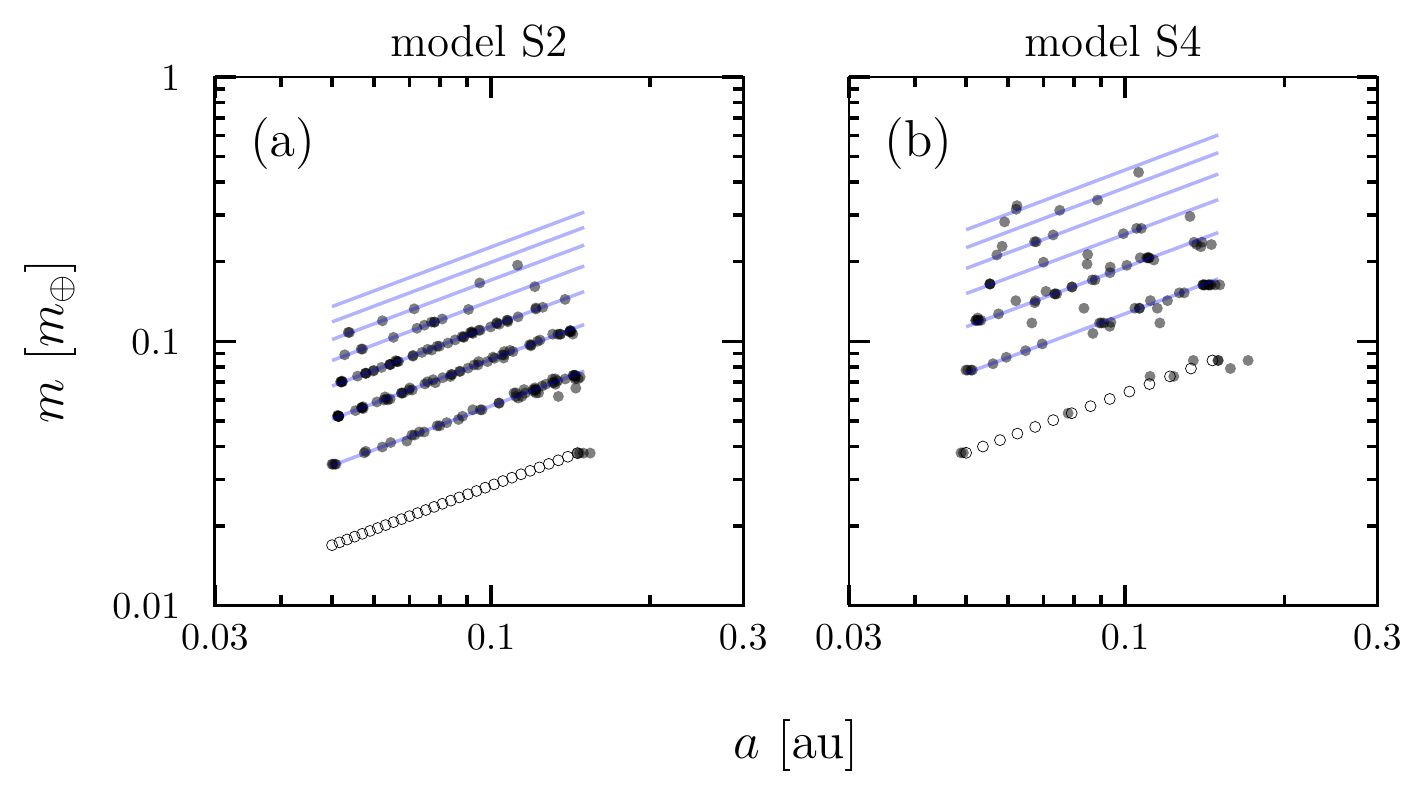}
\caption{Mass of the final planets against the semi-major axis for all runs (filled circles) together with the initial conditions (open circles) for models S2 (a) and S4 (b).
The blue lines show the theoretical predictions at each step of the tournament-like merger.}
\label{fig:a-m_2-4}
\end{figure*}

\begin{figure}
\centering
\includegraphics[width=6.5cm]{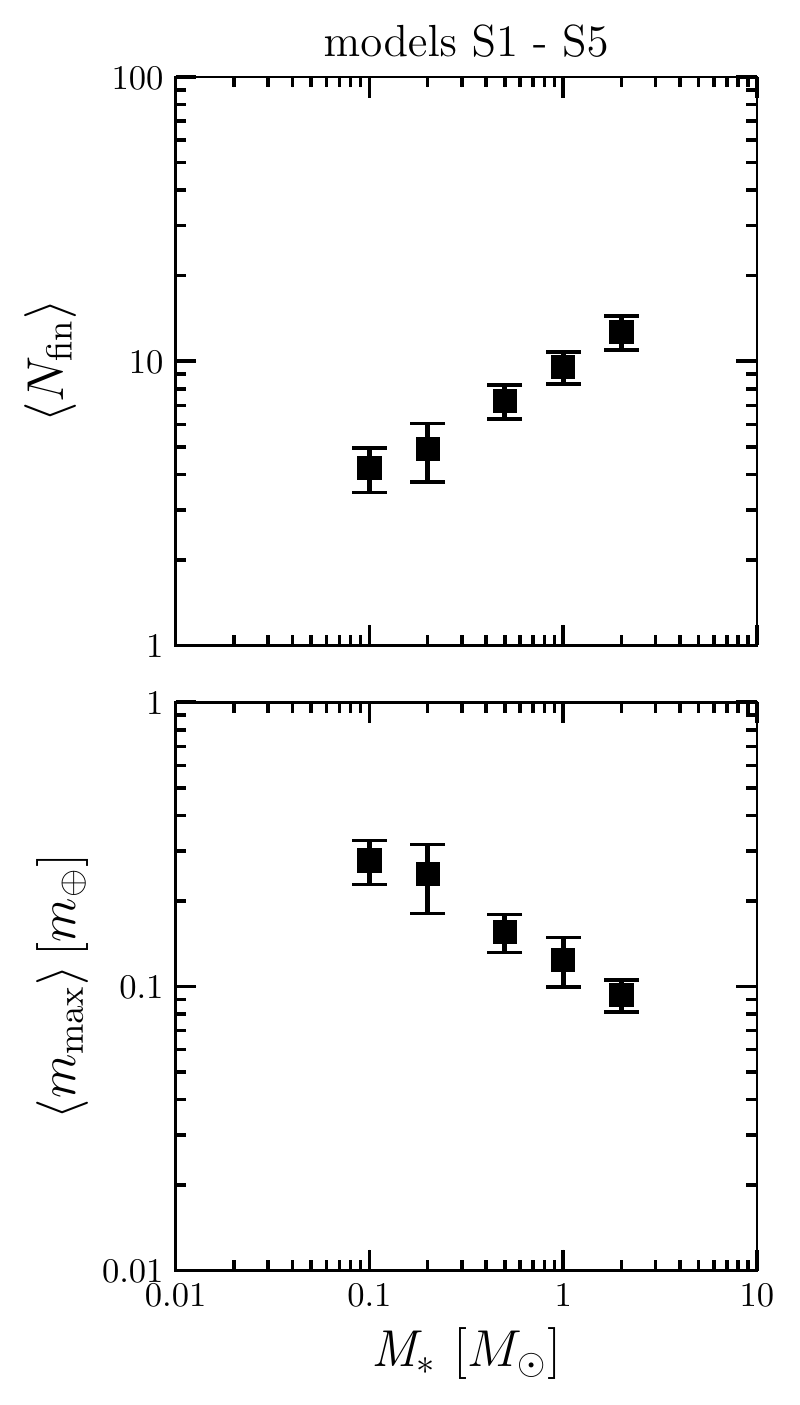}
\caption{Stellar mass dependence of the final planets' properties; average number of planets $\langle N_{\rm{fin}} \rangle$ (top panel) and mass of the heaviest planet $\langle m_{\rm{max}} \rangle$ (bottom panel) for models S1--S5. The error bar indicates the deviation.}
\label{fig:M-N-m1}
\end{figure}

\subsection{Overall Evolution}

We take two models, models S2 (1 $M_{\odot}$) and S4 (0.2 $M_{\odot}$), for a comparison between Sun-like stars and M-type stars, and show the overall picture of the simulations. 
First, Fig. \ref{fig:e-a_snap} presents examples of the evolution of protoplanet systems on the semimajor axis $a$--eccentricity $e$ plane.
Protoplanets perturb each other and collide with one another to form planets. 
In model S2, 30 protoplanets grow to ten planets after $2\times 10^8$ $t_\mathrm{K}$, while in model S4, five planets are formed from 14 protoplanets.
During the evolution the planetary orbits tend to be more eccentric in model S4 than in model S2.
At $t = 2\times 10^8$ $t_\mathrm{K}$ in model S2, the average of the eccentricity of the planet is 0.013 and the orbital separation is 0.0099 au, while in model S4, the eccentricity is 0.033 and the orbital separation is 0.021 au.
It is clear that the stellar mass affects the architecture of planetary systems.

Next, we show the time evolution of the semi-major axis of the same run in Fig. \ref{fig:a-t_2-4}.
The number of planets gradually decreases and the orbital interval becomes wider.
This giant impact process settles down in about $10^7$--$10^8$ $t_\mathrm{K}$.
In panel (a), planets collide and merge mostly with their neighbors.
This evolution is like a tournament chart.
However, panel (b) is characterized by a large final eccentricity of the planet and frequent orbital crossings.
As a result, the final planets have a large eccentricity and a wide spacing.
In both models, no planets deviate from their initial distribution range.

The mass of the final planets are plotted against the semi-major axis for all 20 runs in models S2 and S4 together with the initial mass in Fig. \ref{fig:a-m_2-4}.
The planets are more linearly aligned in panel (a) than panel (b), which suggests that in model S2 the accretion proceeds locally keeping the initial global mass distribution.
We theoretically estimate the possible mass distributions at each giant impact step.
We assume that merging of adjacent planets in a tournament-like way.
In a stage all neighboring planets merge with each other.
After merging, new planets are located at the position of the center of mass of the merging planets.
Planets in the $n$th stage are formed by $n$ collisions and consist of $2^n$ protoplanets.
The growth pattern is like a tournament chart. 
In cases there are an odd number of planets the seeding is incorporated as appropriate.
The theoretical lines for planet masses formed by merging of two to eight protoplanets are shown in both panels.
We find that in panel (a) the distribution of the final planets falls on this line nicely.
This result does not apply to panel (b) since the giant impact takes place more globally as the stellar mass decreases.

\subsection{Stellar Mass Dependence} \label{sec:stellar_mass_dependence}

We now consider the planetary system parameters as a function of the stellar mass, and show the statistical results from 20 runs per model.
Fig. \ref{fig:M-N-m1} shows the number of planets $\langle N_{\rm{fin}} \rangle$ and the maximum mass $\langle m_{\rm{max}} \rangle$ of the final planetary system for models S1--S5.
As the stellar mass increases, $\langle N_{\rm{fin}} \rangle$ increases while $\langle m_{\rm{max}} \rangle$ decreases.
We can describe the dependence as $ \langle N_{\rm{fin}} \rangle \propto \left( M_* / M_\odot  \right) ^{0.379}$.
This result shows that since the total disk mass is fixed, the maximum mass is inversely proportional to the number.
We find that $\langle m_{\rm{max}} \rangle \propto \left( M_* / M_\odot  \right) ^{-0.379}$.

\begin{figure}
\centering
\includegraphics[width=7cm]{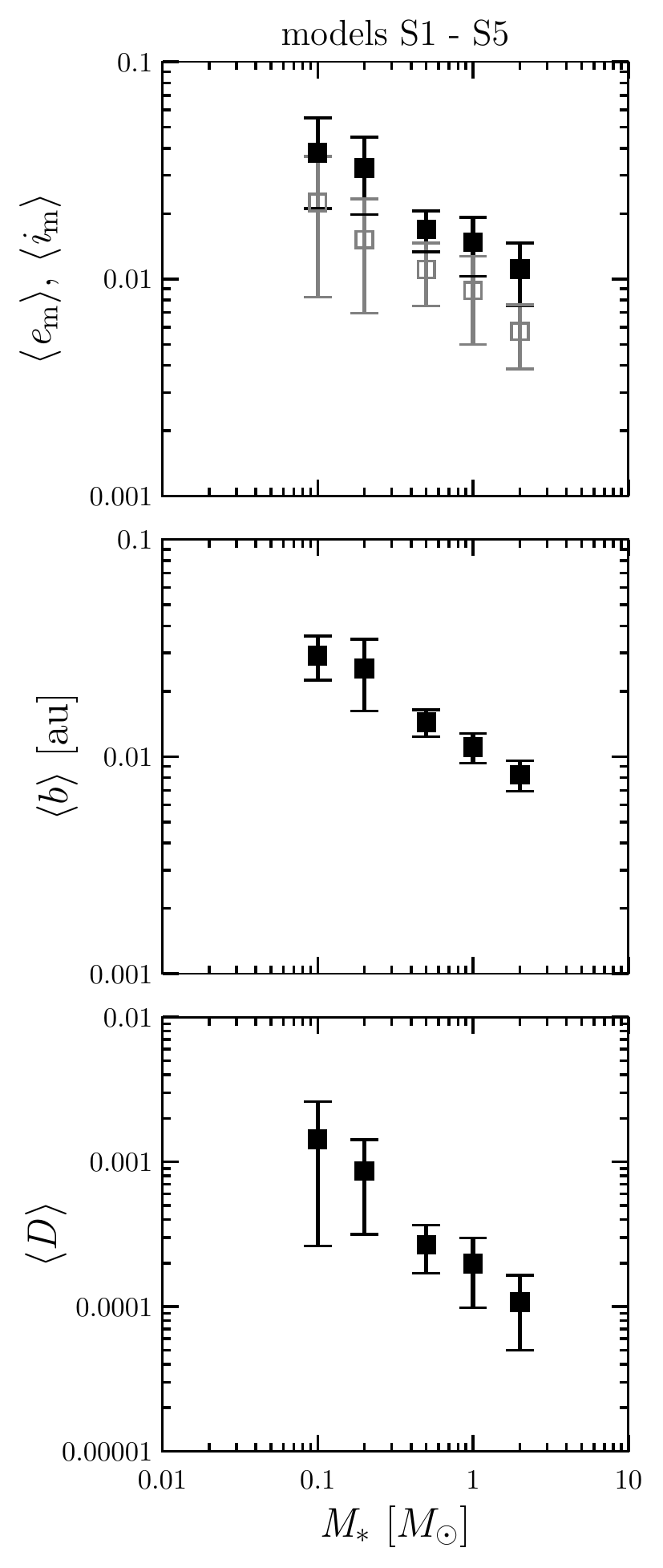} 
\caption{Stellar mass dependence of the mass-weighted eccentricity $\langle e_{\rm{m}} \rangle$, inclination $\langle i_{\rm{m}} \rangle$ (top panel), orbital separation $\langle b \rangle$ (middle panel), and AMD $\langle \it{D} \rangle$ (bottom panel) for models S1--S5.}
\label{fig:M-eibD}
\end{figure}

Fig. \ref{fig:M-eibD} presents the mass dependence of the system parameter.
We find that $\langle e_{\rm{m}} \rangle$, $\langle i_{\rm{m}} \rangle$, $\langle b \rangle$ and $\langle D \rangle$ decrease with increasing $M_*$.
As shown in Eq.(\ref{equ:Hill_radius}),  the Hill radius is proportional to the $-$1/3 power of the stellar mass.
Thus a mass difference of a factor of 1/10 leads to a Hill radius difference of about two-fold.
Around low-mass stars, the gravitational interaction of planets is more effective due to the large Hill radius relative to the physical radius.
As a result, the planet's orbits are more disturbed, in other words, the eccentricity and inclination are larger.
Therefore, the final state of a low-mass star planetary system has a large orbital separation on average. 
We also confirm that the orbital spacing normalized by the mutual Hill radius is within the range of about 19-23 for all stellar masses of models S1-S5.

\subsubsection{Disk Parameter Dependence} \label{sec:sigma_b_dependence}

\begin{figure}
\centering
\includegraphics[width=6.5cm]{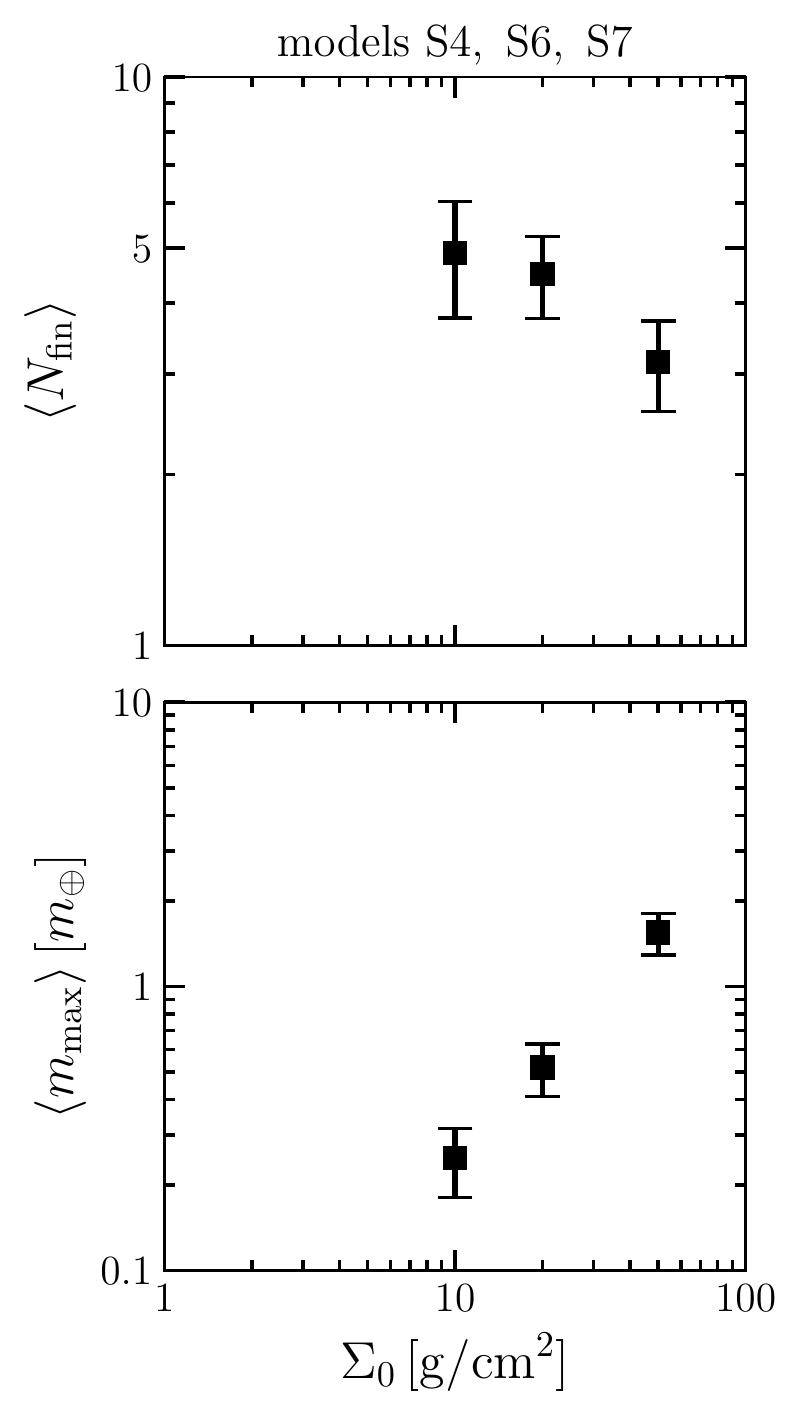} 
\caption{Relationship between the initial disk surface density and the mass of the planet (models S4, S6, and S7). It is clear that the higher the initial disk surface density, the larger the final planet. This dependence is almost linear.}
\label{fig:sigma-n_m}
\end{figure}

\begin{figure}
\centering
\includegraphics[width=6.5cm]{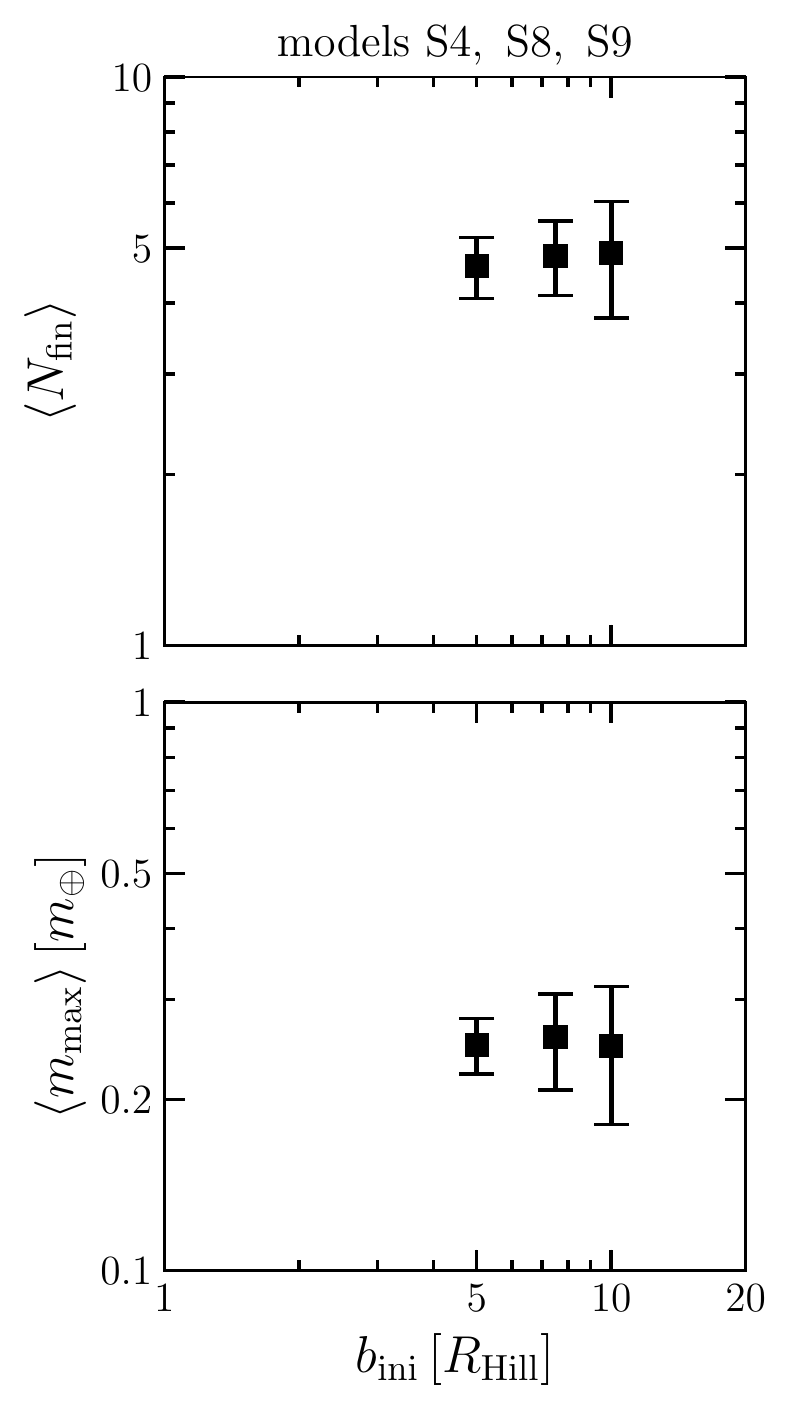}
\caption{Relationship between the number of final planets $\langle N_{\rm{fin}} \rangle$, the maximum mass of planets $\langle m_{\rm{max}} \rangle$, and the initial orbital separation $b_{\rm{ini}}$ (models S4, S8, and S9). }
\label{fig:b-m1}
\end{figure}

We investigate the dependence of the planetary system architecture on the reference surface density $\Sigma_0$ and initial orbital separation of protoplanets $b_{\rm{ini}}$.
We fix the stellar mass as $0.2M_\odot$.
Fig. \ref{fig:sigma-n_m} shows the number $\langle N_{\rm{fin}} \rangle$ and the maximum mass $\langle m_{\rm{max}} \rangle$ of the final planets against $\Sigma_0$ (models S4, S6 and S7).
We find that $\langle m_{\rm{max}} \rangle$ and $\Sigma_0$ have a almost linear relationship, $\langle m_{\rm{max}} \rangle \propto \Sigma_0 ^{1.14}$, which is consistent with the results of \citet[][]{Kokubo_2006}. On the other hand, $\langle N_{\rm{fin}} \rangle$ has a weak negative dependence on $\Sigma_0$.
Fig. \ref{fig:b-m1} shows the relationship between $\langle N_{\rm{fin}} \rangle$, $\langle m_{\rm{max}} \rangle$, and $b_{\rm{ini}}$ (models S4, S8, and S9).
The initial orbital separation does not change the results in the range of 5 to 10 Hill radius. In addition, \citet[][]{Kokubo_2006} showed that the initial orbital separation does not affect the properties of planets in the range of 6 to 12 Hill radius with $M_* = 1M_\odot$.
Our study confirms that this feature holds for low-mass stars.
We also check the influence on the orbital structure.
There are no effects for different $\Sigma_0$ and $b_{\rm{ini}}$ on $e$ and $i$.

\subsubsection{Minimum-Mass Extrasolar Nebula Model}

\begin{figure}
\centering
 \includegraphics[width=6.5cm]{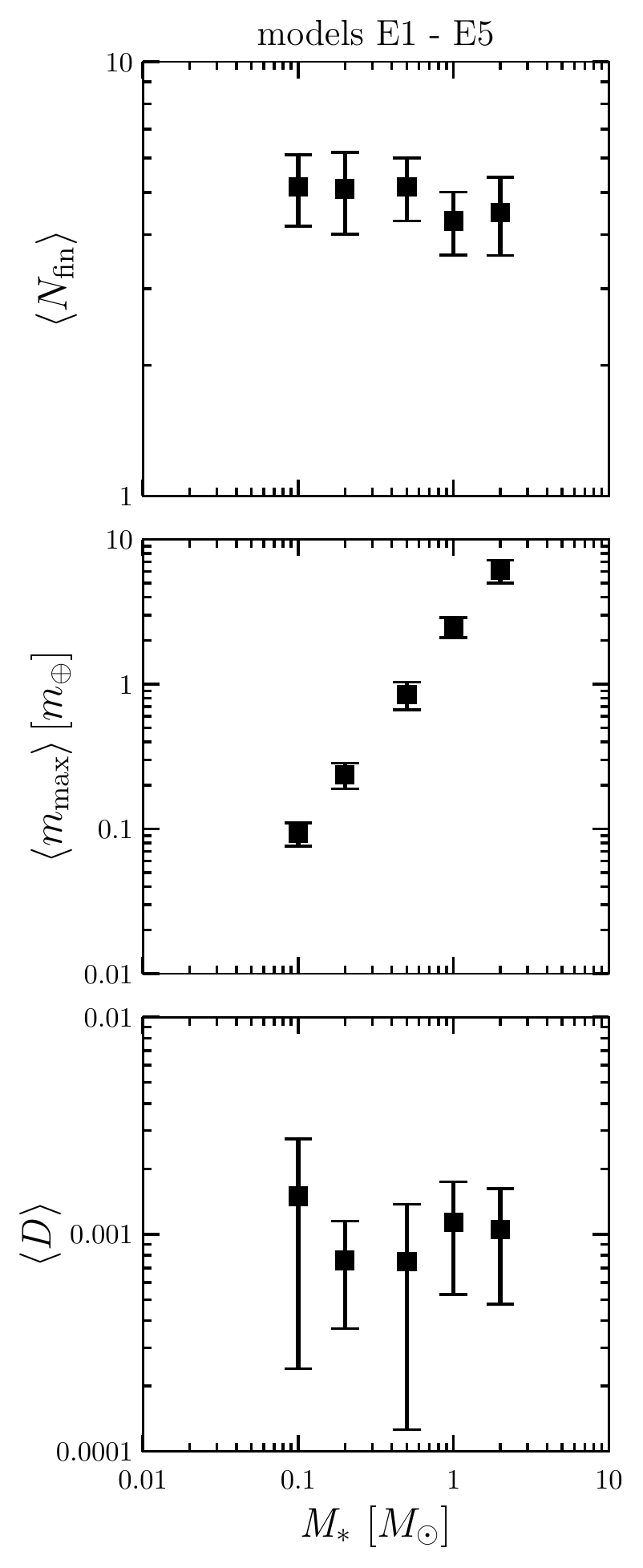} 

\caption{Stellar mass dependence of the number of final planets $\langle N_{\rm{fin}} \rangle$ (top panel), maximum mass $\langle m_{\rm{max}} \rangle$ (middle panel) and averaged AMD $\langle D\rangle$ (bottom panel) for MMEN disk models. The results for the number of planets in the top panel do not change much. In the middle panel, the planet mass is approximately proportional to the stellar mass.}
\label{fig:M-Nm1D_MMEN}
\end{figure}

Using the MMEN model, we investigate how the stellar mass and the disk model affect the structure of planetary systems with fixed orbital radius (models E1--E5).
Fig. \ref{fig:M-Nm1D_MMEN} shows the dependence of $\langle N_{\rm{fin}} \rangle$, $\langle m_{\rm{max}} \rangle$, and $\langle D\rangle$ on $M_*$.
We find that $\langle N_{\rm{fin}} \rangle$ is almost constant.
As shown in \ref{sec:stellar_mass_dependence}, the lower the stellar mass, the more the orbital structure is disturbed and the lower the number of planets.
This effect is weakened by decreasing the surface density or disk mass in the MMEN model.
On the other hand, $\langle m_{\rm{max}} \rangle$ increase with $M_*$, and have a relatively strong dependence, $\langle m_{\rm{max}} \rangle \propto \left( M_* / M_\odot  \right) ^{1.41}$.
As shown in \ref{sec:sigma_b_dependence}, the planet mass increases with the disk surface density.
In the MMEN model, this effect is more prominent than the others, reversing the dependence of $\langle m_{\rm{max}} \rangle$ on $M_*$ in models S1-S5.
The qualitative dependence of $\langle D \rangle$ on $M_*$ seems similar to that of the standard disk model.
As $M_*$ increases, $\langle D \rangle$ becomes smaller.
Compared to models S1--S5,  $\langle D \rangle$ is generally larger. 
This is due to the larger planet masses.

\subsection{Planets in the Habitable Zone}

\begin{figure}
\centering
\includegraphics[width=6.5cm]{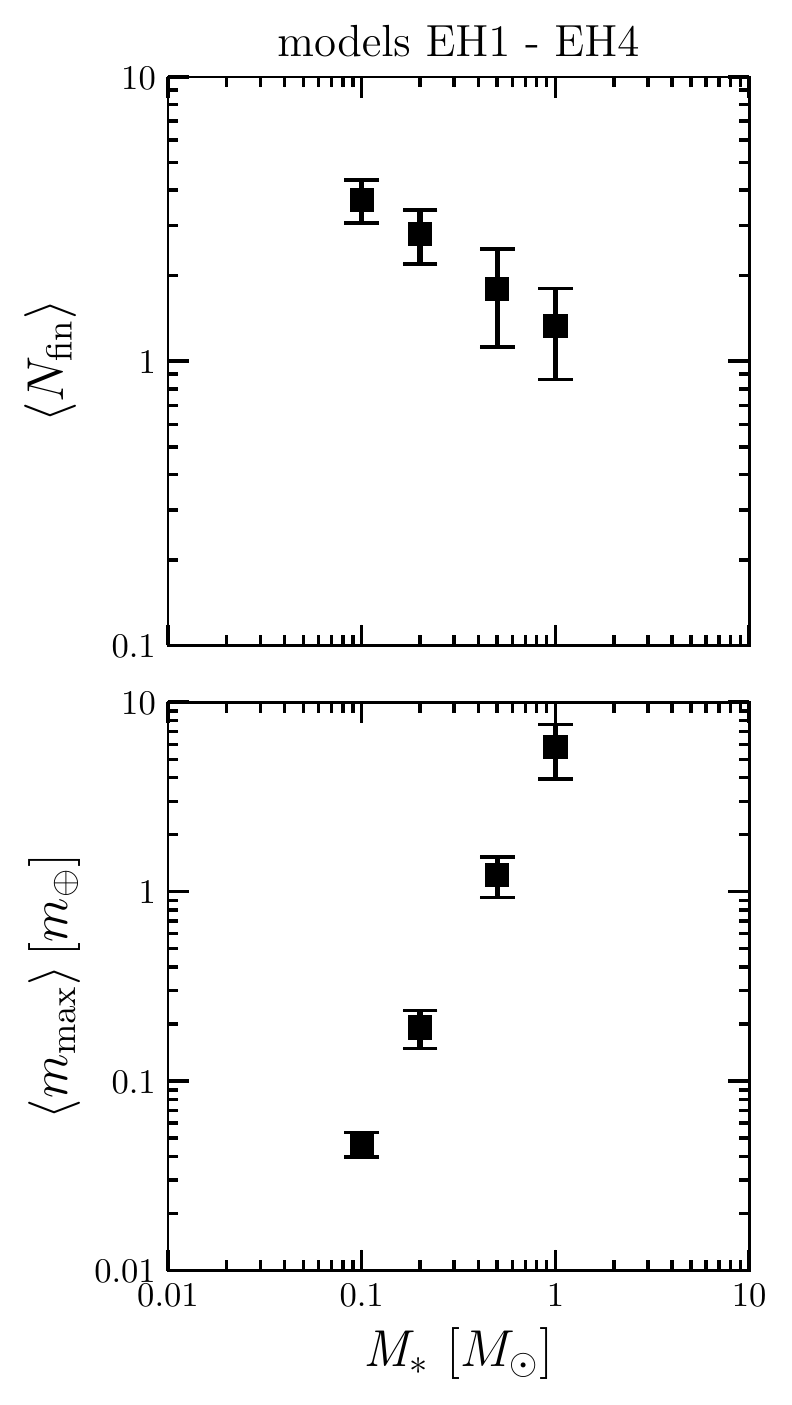} 
\caption{As Fig. \ref{fig:M-N-m1}, but for HZ planets with MMEN (models EH1--EH4).}
\label{fig:M-Nm1_MMEN_HZ}
\end{figure}

We focus on the planets in the HZ and examine their properties using the MMEN model (models EH1--EH4).
Fig. \ref{fig:M-Nm1_MMEN_HZ} shows the dependence of the number and mass of planets on the stellar mass. 
The number of planets $\langle N_{\rm{fin}} \rangle$ decreases as $M_*$ increases, while the maximum mass $\langle m_{\rm{max}} \rangle$ increases with $M_*$.
This trend is consistent with the previous studies \citep[e.g.,][]{Raymond_2007, Ciesla_2015}.
The prediction of the planetary masses in the HZ using our model and Equation (2) of \cite{Raymond_2007} results in $M_{\rm{prediction}} \propto M_*^{1.7}$, which shows a little bit shallower slope than our result $\langle M_{\rm{max}} \rangle \propto M_*^{2.1}$.
In this prediction, the scaling factor of the stellar metallicity is originally included apart from the stellar mass. Our disk model incorporates the metallicity into the stellar mass, which may lead to a stronger dependence.
As we have mentioned, the mass of the planets is directly affected by the disk surface density.
Since the MMEN model is used, it is consistent that the planet mass and the stellar mass are positively correlated.
In some simulation results of model EH1 there are no planets in the HZ.
In the calculation of $\langle N_{\rm{fin}} \rangle$, $N_{\rm{fin}}$=0 cases are included. 
For other values of orbital parameters, we exclude those runs.

\begin{figure}
\centering
\includegraphics[width=7cm]{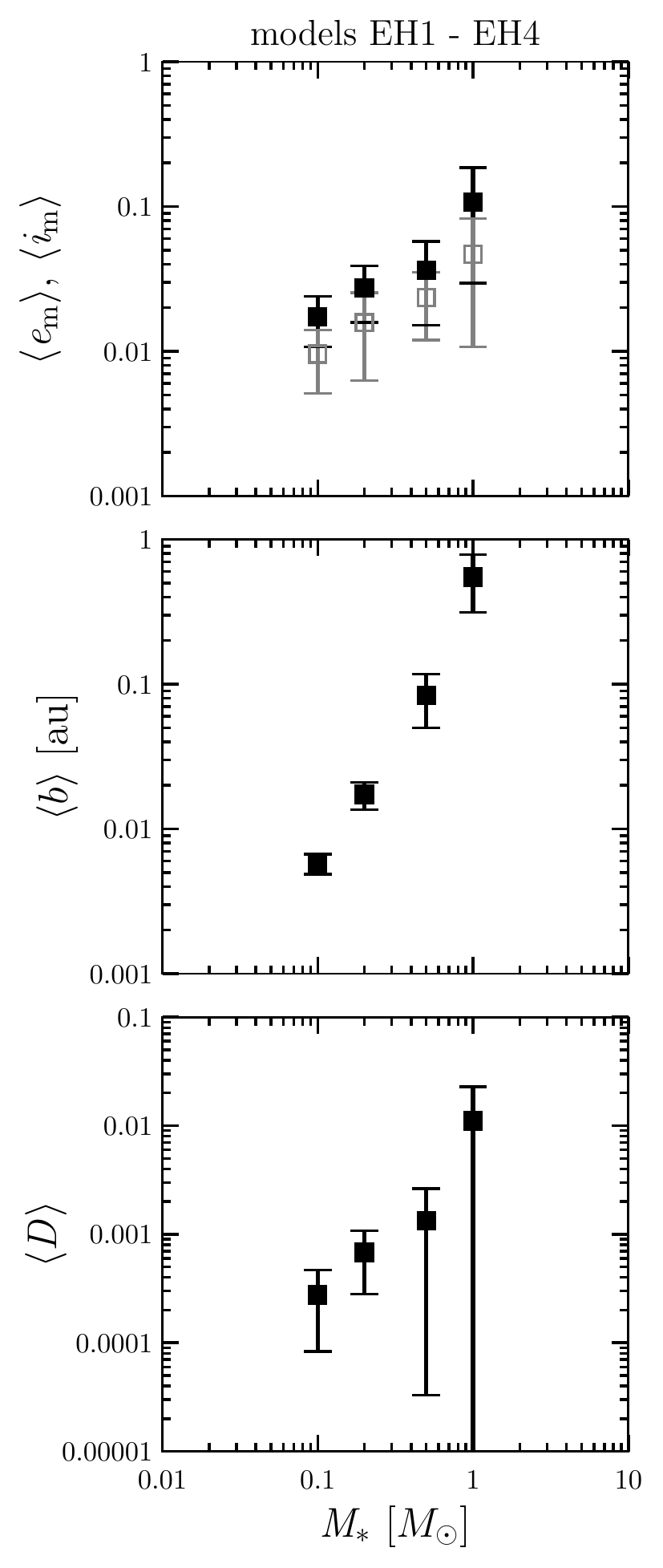} 
\caption{As Fig. \ref{fig:M-eibD}, but for HZ planets with MMEN (models EH1--EH4).}
\label{fig:eibD_MMEN-HZ}
\end{figure}

Fig. \ref{fig:eibD_MMEN-HZ} shows the orbital parameters of planetary systems.
The orbital eccentricity $\langle e_m \rangle$ and inclination $\langle i_m \rangle$ increase with $M_*$ and the orbital separation $\langle b \rangle$ also increases with $M_*$.
When we consider planet formation in the HZ, we need to consider the balance between the effects of strengthening and weakening of gravitational scattering.
As discussed in \ref{sec:stellar_mass_dependence}, with a fixed orbital radius, the Hill radius increases with decreasing stellar mass.
In addition, for smaller stars, the HZ is closer and the Hill radius becomes smaller.
The overall effect of these two factors is that gravitational scattering among protoplanets becomes weaker as the stellar mass decreases, because the distance from the star has a stronger effect.
This can be understood in the following way: the smaller the stellar mass, the weaker the scattering, and thus $\langle e_m \rangle$, $\langle i_m \rangle$, and $\langle D \rangle$ become smaller.

If we know the orbital separation, then we can estimate the number of planets in the HZ.
Based on the results in Fig. \ref{fig:eibD_MMEN-HZ}, if we divide the HZ width by the average orbital spacing, we can estimate the number of planets in the HZ to be 3.63, 2.61, 1.47, and 1.28 for $M_* =$ 0.1, 0.2, 0.5, and 1.0 $M_{\odot}$, respectively.
This estimate shows that the number of planets forming in the HZ decreases as the stellar mass increases.
The number of planets in the HZ calculated from the orbital spacing agrees with the actual number of planets in the simulation results.

Finally, we summarize the results of all models in Table \ref{tab:final}.

\begin{table}
\caption{Final architecture of planetary systems}
\label{tab:final}
\begin{tabular}{ccccccc}
 \hline
  Model & $\langle \it{N_{\rm{fin}}} \rangle$ & $\langle m_{\rm{max}} \rangle $ & $\langle e_{\rm{m}} \rangle$ & $\langle i_{\rm{m}} \rangle$ & $\langle b \rangle$ & $\langle D \rangle$\\
   & & $[m_{\earth}]$ & & & [au] & [$\times 10^{-3}$]\\
 \hline
    S1 & 12.7 & 0.0933 & 0.0111 & 0.00573 & 0.00823 & 0.107 \\
    S2 & 9.55 & 0.124 & 0.0148 & 0.00884 & 0.0111 & 0.199 \\
    S3 & 7.25 & 0.155 & 0.0169 & 0.0111 & 0.0144 & 0.268 \\
    S4 & 4.90 & 0.248 & 0.0324 & 0.0152 & 0.0254 & 0.868 \\
    S5 & 4.20 & 0.277 & 0.0381 & 0.0225 & 0.0292 & 1.43 \\
    S6 & 4.50 & 0.518 & 0.0400 & 0.0190 & 0.0268 & 1.45 \\
    S7 & 6.30 & 1.55 & 0.0620 & 0.0250 & 0.0142 & 3.25 \\
    S8 & 4.85 & 0.257 & 0.0358 & 0.0218 & 0.0261 & 1.20 \\
    S9 & 4.65 & 0.250 & 0.0373 & 0.0287 & 0.0277 & 1.57 \\
 \hline
    E1 & 4.50 & 6.11 & 0.0341 & 0.0190 & 0.0267 & 1.05 \\
    E2 & 4.30 & 2.49 & 0.0377 & 0.0159 & 0.0267 & 1.14 \\
    E3 & 5.15 & 0.850 & 0.0283 & 0.0156 & 0.0242 & 0.750 \\
    E4 & 5.10 & 0.237 & 0.0306 & 0.0147 & 0.0238 & 0.758 \\
    E5 & 5.15 & 0.0931 & 0.0400 & 0.0216 & 0.0256 & 1.50 \\
 \hline
    EH1 & 1.33 & 5.78 & 0.107 & 0.0468 & 0.549 & 11.1 \\
    EH2 & 1.80 & 1.23 & 0.0362 & 0.0235 & 0.0835 & 1.33 \\
    EH3 & 2.80 & 0.192 & 0.0274 & 0.0158 & 0.0173 & 0.680 \\
    EH4 & 3.70 & 0.0467 & 0.0174 & 0.00957 & 0.00577 & 0.276 \\
 \hline
\end{tabular}
 {\raggedright $N_{\rm{fin}}$ is the number of final planets, $m_{\rm{max}}$ is the mass of the heaviest planet, $e_{\rm{m}}$ and $i_{\rm{m}}$ are the mass-weighted eccentricity and inclination, $b$ is the orbital separation, and $D$ is the angular momentum deficit. The "<>" symbol indicates the average value of 20 runs. }
\end{table}

\section{Summary and Discussion}

\begin{figure}
\centering
\includegraphics[width=6.6cm]{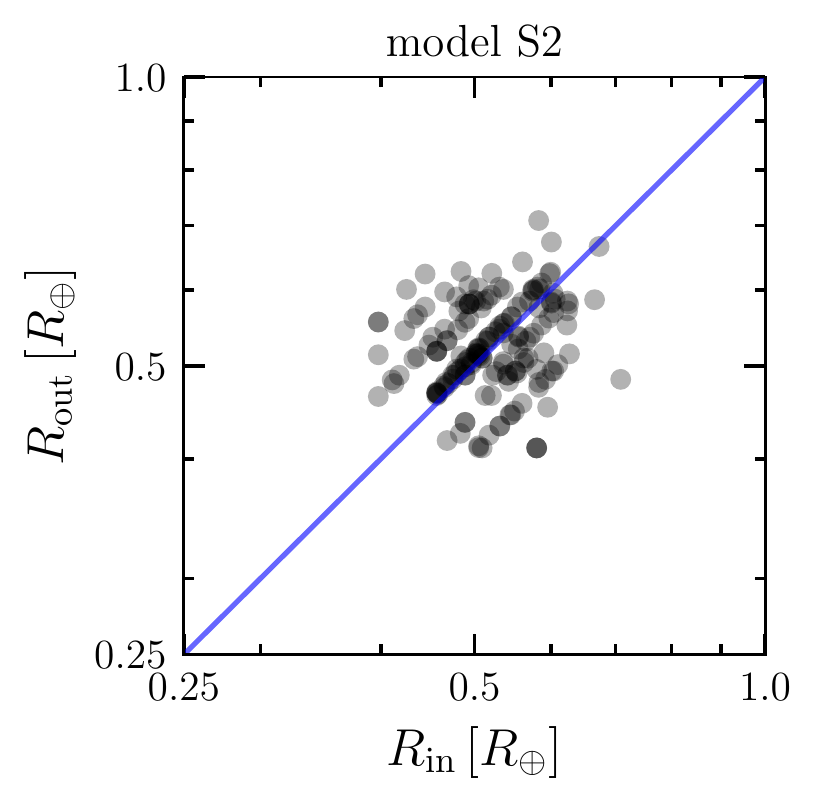} 
\caption{Radius of a planet $R_{\rm{in}}$ versus that of the next planet $R_{\rm{out}}$ for model S2. The blue line indicates $R_{\rm{in}} = R_{\rm{out}}$.}
\label{fig:r_sca}
\end{figure}

\begin{figure}
\centering
\includegraphics[width=6.5cm]{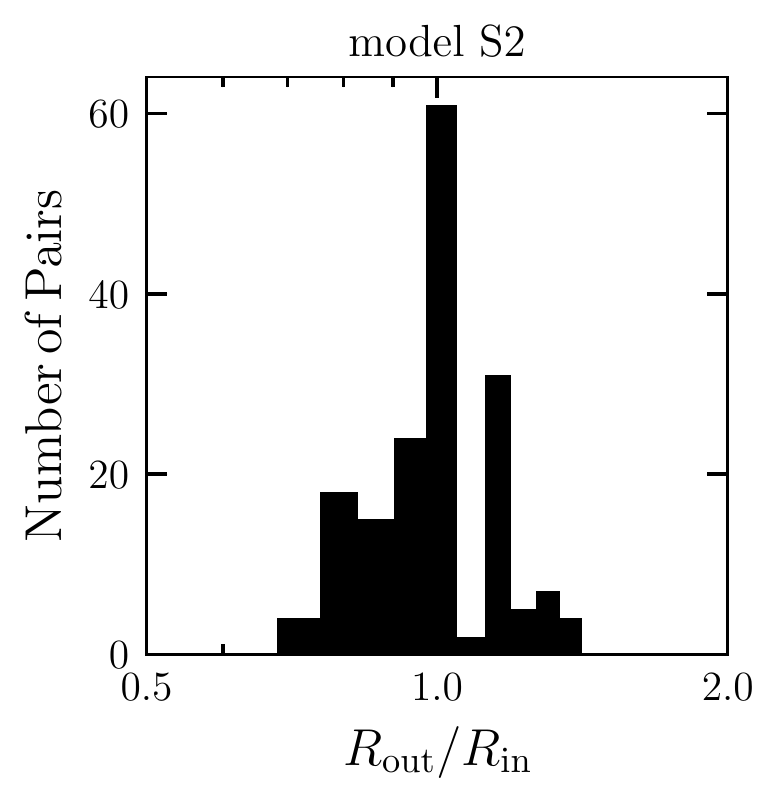} 
\caption{Histogram of the radius ratio of adjacent planet pairs for model S2.}
\label{fig:r_hist}
\end{figure}

\begin{figure}
\centering
\includegraphics[width=6.5cm]{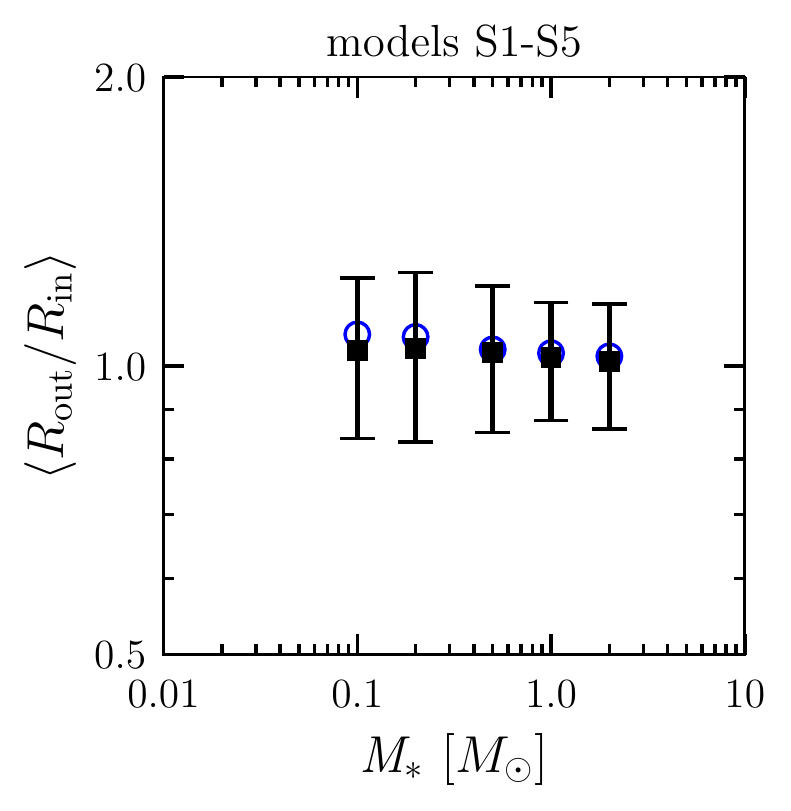} 
\caption{Mean radius ratio of adjacent planet pairs (black squares) plotted against the stellar mass for models S1-S5 with the theoretical predictions (blue circles).
The error bars represent the deviation. }
\label{fig:r_ratio}
\end{figure}

We have performed {\em N}-body simulations of the giant impact stage of terrestrial planet formation changing the stellar mass from 0.1 $M_{\odot}$ to 2 $M_{\odot}$ and examined the effect on the orbital structure of planetary systems.
As the initial conditions, we adopted the isolation mass and distributed protoplanets in 0.05--0.15 au from the central star and the habitable zone (HZ).
We followed the evolution for 200 million orbital periods of the innermost protoplanet.
We investigated the effect of the initial disk parameters and also considered the minimum-mass extrasolar nebula (MMEN)  model.
Our findings are summarized as follows:

\begin{itemize}
\item For a fixed orbital range and mass of the initial protoplanet distribution, the number of planets $\langle N_{\rm{fin}} \rangle$ increases with the stellar mass $M_*$, while the maximum mass of the planets $\langle m_\mathrm{max} \rangle$ decreases. The eccentricity $\langle e_\mathrm{m} \rangle$, inclination $\langle i_\mathrm{m} \rangle$, and orbital separation $\langle b \rangle$ of planet orbits decrease with increasing $M_*$. This is because gravitational interactions among protoplanets become relatively stronger as $M_*$ decreases. 
\item As the total mass of protoplanets increases, $\langle N_{\rm{fin}} \rangle$ decreases, while $\langle m_\mathrm{max} \rangle$ increases.
\item The initial orbital separation of protoplanet systems in the range of 5 to 10 Hill radius does not affect the final orbital structures.
\item In the MMEN model, $\langle m_\mathrm{max} \rangle$ increases with $M_*$.
\item In the HZ of the MMEN model,  $\langle N_{\rm{fin}} \rangle$ decreases with $M_*$, while $\langle m_\mathrm{max} \rangle$ increases.
Other orbital parameters $\langle e_\mathrm{m} \rangle$, $\langle i_\mathrm{m} \rangle$, and $\langle b \rangle$ increase with $M_*$.
\end{itemize}

In this study, the surface density slope $\alpha$ is fixed.
The dependence of $\alpha$ has been discussed in many papers in the case of one solar mass \citep[e.g.,][]{Raymond_2005, Kokubo_2006, Ronco_2014, Izidoro_2015}. 
We confirmed that the stellar mass dependence remains the same for different $\alpha$. 
Details will be discussed in the next paper.

Our results show that orbital separations of planets are 17-26 Hill radii for all models.
This is consistent with the Kepler planets that are about 20 Hill radii apart \citep[e.g.,][]{Weiss_2018}.
We can also reproduce planets in the smaller mass side of the CKS samples \citep[][]{Petigura_2017, Johnson_2017} used in the MMEN model.
However, since there are still few planets around low-mass stars,  our results below one solar mass are rather predictive.
Our concern is that planet masses are inferred from radii in the construction of the MMEN models.
The mass-radius relation used in \citet{Dai_2020} does not yet reflect the latest exoplanet observations.
In addition, when only planets with known masses by the radial velocity and transit timing variation methods are used, the disk surface density is found to be about twice as large.
The absolute values of disk surface densities are still quite uncertain.
Considering this shortage of data and the uncertainty of the mass-radius relation, it is necessary to reconsider the MMEN model, at least for planets around low-mass stars.

It is meaningful to investigate whether the characteristics of planetary systems observed around 1 $M_{\odot}$ stars are also observed around low-mass stars. 
Here we focus on the radius ratio of neighboring planet pairs.
Though we have 20 planetary systems per model, we do not consider which system they belong to but treat all of them equally as a planet pair.
Fig. \ref{fig:r_sca} shows the correlation between the radii of the inner and outer planets for model S2. 
Neighboring planets tend to have the same size. 
In addition, we plot the number of planet pairs against the radius ratio of the outer to inner planets in Fig. \ref{fig:r_hist}.
We find that the radius ratio distribution has a peak around 1.

We show the radius ratio of neighboring pairs against the stellar mass in Fig. \ref{fig:r_ratio} (models S1-S5) together with the theoretical prediction. 
The theoretical prediction is based on the tournament-like growth discussed in Section 3.1. 
We also used the number of planets and orbital separation obtained in the simulations. 
We find that the radius ratio of neighboring planets does not depend on the stellar mass.
These analyses conclude that the neighboring planets have similar radii in planetary systems around $\sim 0.1$ au if the initial protoplanet mass does not strongly depend on the orbital radius,  which is consistent with the peas-in-a-pod pattern seen in the Kepler multiple-planet system \citep[e.g.,][]{Weiss_2018}.

In this study, we mainly focused on the dependence of the orbital structure of planetary systems on the stellar mass. 
In reality, the orbital structure may depend on disk and protoplanet parameters that are not considered here.
A comparison using dimensionless orbital parameters such as those normalized by the Hill radius is helpful for physical understanding of the structure dependence.
By performing a systematic parameter survey we will generalize the orbital parameter dependence and discuss a new basic scaling law of the structure that incorporates not only gravitational interaction but also collisions among planets in the next paper.

\section*{Acknowledgements}
E.Kokubo is supported by JSPS KAKENHI grant No. 18H05438.

\section*{Data Availability} 
All data underlying this research are included in the article.

\bibliographystyle{mnras}
\bibliography{Hoshino_Kokubo} 

\begin{thebibliography}{}
\makeatletter
\relax
\def\mn@urlcharsother{\let\do\@makeother \do\$\do\&\do\#\do\^\do\_\do\%\do\~}
\def\mn@doi{\begingroup\mn@urlcharsother \@ifnextchar [ {\mn@doi@}
  {\mn@doi@[]}}
\def\mn@doi@[#1]#2{\def\@tempa{#1}\ifx\@tempa\@empty \href
  {http://dx.doi.org/#2} {doi:#2}\else \href {http://dx.doi.org/#2} {#1}\fi
  \endgroup}
\def\mn@eprint#1#2{\mn@eprint@#1:#2::\@nil}
\def\mn@eprint@arXiv#1{\href {http://arxiv.org/abs/#1} {{\tt arXiv:#1}}}
\def\mn@eprint@dblp#1{\href {http://dblp.uni-trier.de/rec/bibtex/#1.xml}
  {dblp:#1}}
\def\mn@eprint@#1:#2:#3:#4\@nil{\def\@tempa {#1}\def\@tempb {#2}\def\@tempc
  {#3}\ifx \@tempc \@empty \let \@tempc \@tempb \let \@tempb \@tempa \fi \ifx
  \@tempb \@empty \def\@tempb {arXiv}\fi \@ifundefined
  {mn@eprint@\@tempb}{\@tempb:\@tempc}{\expandafter \expandafter \csname
  mn@eprint@\@tempb\endcsname \expandafter{\@tempc}}}

\bibitem[\protect\citeauthoryear{Ansdell, Williams, Manara, Miotello, Facchini,
  van~der Marel, Testi  \& van Dishoeck}{Ansdell et~al.}{2017}]{Ansdell_2017}
Ansdell M.,  Williams J.~P.,  Manara C.~F.,  Miotello A.,  Facchini S.,
  van~der Marel N.,  Testi L.,   van Dishoeck E.~F.,  2017, \mn@doi [\aj]
  {10.3847/1538-3881/aa69c0}, 153, 240

\bibitem[\protect\citeauthoryear{Bochanski, Hawley, Covey, West, Reid,
  Golimowski  \& Ivezi{\'{c}}}{Bochanski et~al.}{2010}]{Bochanski_2010}
Bochanski J.~J.,  Hawley S.~L.,  Covey K.~R.,  West A.~A.,  Reid I.~N.,
  Golimowski D.~A.,   Ivezi{\'{c}} {\v{Z}}.,  2010, \mn@doi [\aj]
  {10.1088/0004-6256/139/6/2679}, 139, 2679

\bibitem[\protect\citeauthoryear{Chiang \& Laughlin}{Chiang \&
  Laughlin}{2013}]{Chiang_2013}
Chiang E.,  Laughlin G.,  2013, \mn@doi [\mnras] {10.1093/mnras/stt424}, 431,
  3444

\bibitem[\protect\citeauthoryear{Ciesla, Mulders, Pascucci  \& Apai}{Ciesla
  et~al.}{2015}]{Ciesla_2015}
Ciesla F.~J.,  Mulders G.~D.,  Pascucci I.,   Apai D.,  2015, \mn@doi [\apj]
  {10.1088/0004-637x/804/1/9}, 804, 9

\bibitem[\protect\citeauthoryear{Dai, Winn, Schlaufman, Wang, Weiss, Petigura,
  Howard  \& Fang}{Dai et~al.}{2020}]{Dai_2020}
Dai F.,  Winn J.~N.,  Schlaufman K.,  Wang S.,  Weiss L.,  Petigura E.~A.,
  Howard A.~W.,   Fang M.,  2020, \mn@doi [\aj] {10.3847/1538-3881/ab88b8},
  159, 247

\bibitem[\protect\citeauthoryear{Dressing \& Charbonneau}{Dressing \&
  Charbonneau}{2013}]{Dressing_2013}
Dressing C.~D.,  Charbonneau D.,  2013, \mn@doi [\apj]
  {10.1088/0004-637x/767/1/95}, 767, 95

\bibitem[\protect\citeauthoryear{Dressing \& Charbonneau}{Dressing \&
  Charbonneau}{2015}]{Dressing_2015}
Dressing C.~D.,  Charbonneau D.,  2015, \mn@doi [\apj]
  {10.1088/0004-637x/807/1/45}, 807, 45

\bibitem[\protect\citeauthoryear{{Esteves}, {Izidoro}, {Bitsch}, {Jacobson},
  {Raymond}, {Deienno}  \& {Winter}}{{Esteves} et~al.}{2022}]{Esteves_2022}
{Esteves} L.,  {Izidoro} A.,  {Bitsch} B.,  {Jacobson} S.~A.,  {Raymond} S.~N.,
   {Deienno} R.,   {Winter} O.~C.,  2022, \mn@doi [\mnras]
  {10.1093/mnras/stab3203}, \href
  {https://ui.adsabs.harvard.edu/abs/2022MNRAS.509.2856E} {509, 2856}

\bibitem[\protect\citeauthoryear{Harakawa et~al.,}{Harakawa
  et~al.}{2022}]{Harakawa_2022}
Harakawa H.,  et~al., 2022, \mn@doi [\pasj] {10.1093/pasj/psac044}, 74, 904

\bibitem[\protect\citeauthoryear{Hardegree-Ullman, Cushing, Muirhead  \&
  Christiansen}{Hardegree-Ullman et~al.}{2019}]{Hardegree-Ullman_2019}
Hardegree-Ullman K.~K.,  Cushing M.~C.,  Muirhead P.~S.,   Christiansen J.~L.,
  2019, \mn@doi [\aj] {10.3847/1538-3881/ab21d2}, 158, 75

\bibitem[\protect\citeauthoryear{Hartmann \& Davis}{Hartmann \&
  Davis}{1975}]{Hartmann_1975}
Hartmann W.~K.,  Davis D.~R.,  1975, \mn@doi [\icarus]
  {https://doi.org/10.1016/0019-1035(75)90070-6}, 24, 504

\bibitem[\protect\citeauthoryear{{Hayashi}}{{Hayashi}}{1981}]{Hayashi_1981}
{Hayashi} C.,  1981, \mn@doi [Progress of Theoretical Physics Supplement]
  {10.1143/PTPS.70.35}, \href
  {https://ui.adsabs.harvard.edu/abs/1981PThPS..70...35H} {70, 35}

\bibitem[\protect\citeauthoryear{{Hayashi}, {Nakazawa}  \&
  {Nakagawa}}{{Hayashi} et~al.}{1985}]{Hayashi_1985}
{Hayashi} C.,  {Nakazawa} K.,   {Nakagawa} Y.,  1985, in {Black} D.~C.,
  {Matthews} M.~S.,  eds, Protostars and Planets II. pp 1100--1153

\bibitem[\protect\citeauthoryear{{He}, {Ford}  \& {Ragozzine}}{{He}
  et~al.}{2021}]{He_2021}
{He} M.~Y.,  {Ford} E.~B.,   {Ragozzine} D.,  2021, \mn@doi [\aj]
  {10.3847/1538-3881/abc68b}, \href
  {https://ui.adsabs.harvard.edu/abs/2021AJ....161...16H} {161, 16}

\bibitem[\protect\citeauthoryear{{Izidoro}, {Raymond}, {Morbidelli}  \&
  {Winter}}{{Izidoro} et~al.}{2015}]{Izidoro_2015}
{Izidoro} A.,  {Raymond} S.~N.,  {Morbidelli} A.,   {Winter} O.~C.,  2015,
  \mn@doi [\mnras] {10.1093/mnras/stv1835}, \href
  {https://ui.adsabs.harvard.edu/abs/2015MNRAS.453.3619I} {453, 3619}

\bibitem[\protect\citeauthoryear{{Johnson} et~al.,}{{Johnson}
  et~al.}{2017}]{Johnson_2017}
{Johnson} J.~A.,  et~al., 2017, \mn@doi [\aj] {10.3847/1538-3881/aa80e7}, \href
  {https://ui.adsabs.harvard.edu/abs/2017AJ....154..108J} {154, 108}

\bibitem[\protect\citeauthoryear{{Kaminski, A.} et~al.,}{{Kaminski, A.}
  et~al.}{2018}]{Kaminski_2018}
{Kaminski, A.} et~al., 2018, \mn@doi [\aap] {10.1051/0004-6361/201833354}, 618,
  A115

\bibitem[\protect\citeauthoryear{{Kokubo} \& {Genda}}{{Kokubo} \&
  {Genda}}{2010}]{Kokubo_2010}
{Kokubo} E.,  {Genda} H.,  2010, \mn@doi [\apjl] {10.1088/2041-8205/714/1/L21},
  \href {https://ui.adsabs.harvard.edu/abs/2010ApJ...714L..21K} {714, L21}

\bibitem[\protect\citeauthoryear{Kokubo \& Ida}{Kokubo \&
  Ida}{1998}]{Kokubo_1998}
Kokubo E.,  Ida S.,  1998, \mn@doi [\icarus]
  {https://doi.org/10.1006/icar.1997.5840}, 131, 171

\bibitem[\protect\citeauthoryear{Kokubo \& Ida}{Kokubo \&
  Ida}{2000}]{Kokubo_2000}
Kokubo E.,  Ida S.,  2000, \mn@doi [\icarus]
  {https://doi.org/10.1006/icar.1999.6237}, 143, 15

\bibitem[\protect\citeauthoryear{Kokubo \& Ida}{Kokubo \&
  Ida}{2002}]{Kokubo_2002}
Kokubo E.,  Ida S.,  2002, \mn@doi [\apj] {10.1086/344105}, 581, 666

\bibitem[\protect\citeauthoryear{Kokubo \& Ida}{Kokubo \&
  Ida}{2012}]{Kokubo_2012}
Kokubo E.,  Ida S.,  2012, \mn@doi [Prog. Theor. Exp. Phys.]
  {10.1093/ptep/pts032}, 2012

\bibitem[\protect\citeauthoryear{Kokubo \& Makino}{Kokubo \&
  Makino}{2004}]{Kokubo_2004}
Kokubo E.,  Makino J.,  2004, \mn@doi [\pasj] {10.1093/pasj/56.5.861}, 56, 861

\bibitem[\protect\citeauthoryear{{Kokubo}, {Yoshinaga}  \& {Makino}}{{Kokubo}
  et~al.}{1998}]{Kokubo_1998_MNRAS}
{Kokubo} E.,  {Yoshinaga} K.,   {Makino} J.,  1998, \mn@doi [\mnras]
  {10.1046/j.1365-8711.1998.01581.x}, \href
  {https://ui.adsabs.harvard.edu/abs/1998MNRAS.297.1067K} {297, 1067}

\bibitem[\protect\citeauthoryear{Kokubo, Kominami  \& Ida}{Kokubo
  et~al.}{2006}]{Kokubo_2006}
Kokubo E.,  Kominami J.,   Ida S.,  2006, \mn@doi [\apj] {10.1086/501448}, 642,
  1131

\bibitem[\protect\citeauthoryear{{Lambrechts} \& {Johansen}}{{Lambrechts} \&
  {Johansen}}{2012}]{Lambrechts-Johansen_2012}
{Lambrechts} M.,  {Johansen} A.,  2012, \mn@doi [\aap]
  {10.1051/0004-6361/201219127}, \href
  {https://ui.adsabs.harvard.edu/abs/2012A&A...544A..32L} {544, A32}

\bibitem[\protect\citeauthoryear{{Laskar}}{{Laskar}}{1997}]{Laskar_1997}
{Laskar} J.,  1997, \aap, \href
  {https://ui.adsabs.harvard.edu/abs/1997A&A...317L..75L} {317, L75}

\bibitem[\protect\citeauthoryear{{Mahadevan}, {Ramsey}, {Wolszczan}, {Wright},
  {Endl}  \& {Redman}}{{Mahadevan} et~al.}{2010}]{Mahadevan_2010}
{Mahadevan} S.,  {Ramsey} L.,  {Wolszczan} A.,  {Wright} J.,  {Endl} M.,
  {Redman} S.,  2010, in American Astronomical Society Meeting Abstracts \#215.
  p. 421.23

\bibitem[\protect\citeauthoryear{{Makino}}{{Makino}}{1991}]{Makino_1991}
{Makino} J.,  1991, \pasj, \href
  {https://ui.adsabs.harvard.edu/abs/1991PASJ...43..859M} {43, 859}

\bibitem[\protect\citeauthoryear{{Makino} \& {Aarseth}}{{Makino} \&
  {Aarseth}}{1992}]{Makino_1992}
{Makino} J.,  {Aarseth} S.~J.,  1992, \pasj, \href
  {https://ui.adsabs.harvard.edu/abs/1992PASJ...44..141M} {44, 141}

\bibitem[\protect\citeauthoryear{{Micheau} et~al.,}{{Micheau}
  et~al.}{2012}]{Micheau_2012}
{Micheau} Y.,  et~al., 2012, in {McLean} I.~S.,  {Ramsay} S.~K.,   {Takami} H.,
   eds,  Society of Photo-Optical Instrumentation Engineers (SPIE) Conference
  Series Vol. 8446, Ground-based and Airborne Instrumentation for Astronomy IV.
  p. 84462R

\bibitem[\protect\citeauthoryear{Moriarty \& Ballard}{Moriarty \&
  Ballard}{2016}]{Moriarty_2016}
Moriarty J.,  Ballard S.,  2016, \mn@doi [\apj] {10.3847/0004-637x/832/1/34},
  832, 34

\bibitem[\protect\citeauthoryear{{Mulders}, {Dr{\k{a}}{\.z}kowska}, {van der
  Marel}, {Ciesla}  \& {Pascucci}}{{Mulders} et~al.}{2021}]{Mulders_2021}
{Mulders} G.~D.,  {Dr{\k{a}}{\.z}kowska} J.,  {van der Marel} N.,  {Ciesla}
  F.~J.,   {Pascucci} I.,  2021, \mn@doi [\apjl] {10.3847/2041-8213/ac2947},
  \href {https://ui.adsabs.harvard.edu/abs/2021ApJ...920L...1M} {920, L1}

\bibitem[\protect\citeauthoryear{{Nakazawa} \& {Ida}}{{Nakazawa} \&
  {Ida}}{1988}]{Nakazawa_1988}
{Nakazawa} K.,  {Ida} S.,  1988, \mn@doi [Prog. Theor. Phys. Suppl.]
  {10.1143/PTPS.96.167}, \href
  {https://ui.adsabs.harvard.edu/abs/1988PThPS..96..167N} {96, 167}

\bibitem[\protect\citeauthoryear{{Ormel} \& {Klahr}}{{Ormel} \&
  {Klahr}}{2010}]{Ormel-Klahr_2010}
{Ormel} C.~W.,  {Klahr} H.~H.,  2010, \mn@doi [\aap]
  {10.1051/0004-6361/201014903}, \href
  {https://ui.adsabs.harvard.edu/abs/2010A&A...520A..43O} {520, A43}

\bibitem[\protect\citeauthoryear{Pascucci, Mulders, Gould  \&
  Fernandes}{Pascucci et~al.}{2018}]{Pascucci_2018}
Pascucci I.,  Mulders G.~D.,  Gould A.,   Fernandes R.,  2018, \mn@doi [\apj]
  {10.3847/2041-8213/aab6ac}, 856, L28

\bibitem[\protect\citeauthoryear{{Petigura} et~al.,}{{Petigura}
  et~al.}{2017}]{Petigura_2017}
{Petigura} E.~A.,  et~al., 2017, \mn@doi [\aj] {10.3847/1538-3881/aa80de},
  \href {https://ui.adsabs.harvard.edu/abs/2017AJ....154..107P} {154, 107}

\bibitem[\protect\citeauthoryear{{Quirrenbach} et~al.,}{{Quirrenbach}
  et~al.}{2010}]{Quirrenbach_2010}
{Quirrenbach} A.,  et~al., 2010, in {McLean} I.~S.,  {Ramsay} S.~K.,   {Takami}
  H.,  eds,  Society of Photo-Optical Instrumentation Engineers (SPIE)
  Conference Series Vol. 7735, Ground-based and Airborne Instrumentation for
  Astronomy III. p. 773513

\bibitem[\protect\citeauthoryear{{Raymond}, {Quinn}  \& {Lunine}}{{Raymond}
  et~al.}{2005}]{Raymond_2005}
{Raymond} S.~N.,  {Quinn} T.,   {Lunine} J.~I.,  2005, \mn@doi [\apj]
  {10.1086/433179}, \href
  {https://ui.adsabs.harvard.edu/abs/2005ApJ...632..670R} {632, 670}

\bibitem[\protect\citeauthoryear{Raymond, Scalo  \& Meadows}{Raymond
  et~al.}{2007}]{Raymond_2007}
Raymond S.~N.,  Scalo J.,   Meadows V.~S.,  2007, \mn@doi [\apj]
  {10.1086/521587}, 669, 606

\bibitem[\protect\citeauthoryear{{Raymond}, {Kokubo}, {Morbidelli}, {Morishima}
   \& {Walsh}}{{Raymond} et~al.}{2014}]{Raymond_2014}
{Raymond} S.~N.,  {Kokubo} E.,  {Morbidelli} A.,  {Morishima} R.,   {Walsh}
  K.~J.,  2014, in {Beuther} H.,  {Klessen} R.~S.,  {Dullemond} C.~P.,
  {Henning} T.,  eds, Protostars and Planets VI. p.~595

\bibitem[\protect\citeauthoryear{{Ronco} \& {de El{\'\i}a}}{{Ronco} \& {de
  El{\'\i}a}}{2014}]{Ronco_2014}
{Ronco} M.~P.,  {de El{\'\i}a} G.~C.,  2014, \mn@doi [\aap]
  {10.1051/0004-6361/201323313}, \href
  {https://ui.adsabs.harvard.edu/abs/2014A&A...567A..54R} {567, A54}

\bibitem[\protect\citeauthoryear{Scalo et~al.,}{Scalo
  et~al.}{2007}]{Scalo_2007}
Scalo J.,  et~al., 2007, \mn@doi [Astrobiology] {10.1089/ast.2006.0125}, 7, 85

\bibitem[\protect\citeauthoryear{Tamura et~al.,}{Tamura
  et~al.}{2012}]{Tamura_2012}
Tamura M.,  et~al., 2012, \mn@doi [Ground-based and Airborne Instrumentation
  for Astronomy IV] {10.1117/12.925885}, 8446, 638

\bibitem[\protect\citeauthoryear{{Wallace}, {Tremaine}  \&
  {Chambers}}{{Wallace} et~al.}{2017}]{Wallace_2017}
{Wallace} J.,  {Tremaine} S.,   {Chambers} J.,  2017, \mn@doi [\aj]
  {10.3847/1538-3881/aa8c08}, \href
  {https://ui.adsabs.harvard.edu/abs/2017AJ....154..175W} {154, 175}

\bibitem[\protect\citeauthoryear{Weiss et~al.,}{Weiss
  et~al.}{2018}]{Weiss_2018}
Weiss L.~M.,  et~al., 2018, \mn@doi [\aj] {10.3847/1538-3881/aa9ff6}, 155, 48

\bibitem[\protect\citeauthoryear{Wetherill}{Wetherill}{1990}]{Wetherill_1990}
Wetherill G.~W.,  1990, \mn@doi [Annu. Rev. Earth Planet. Sci.]
  {10.1146/annurev.ea.18.050190.001225}, 18, 205

\bibitem[\protect\citeauthoryear{{Wetherill}}{{Wetherill}}{1996}]{Wetherill_1996}
{Wetherill} G.~W.,  1996, \mn@doi [\icarus] {10.1006/icar.1996.0015}, \href
  {https://ui.adsabs.harvard.edu/abs/1996Icar..119..219W} {119, 219}

\bibitem[\protect\citeauthoryear{{Winn} \& {Fabrycky}}{{Winn} \&
  {Fabrycky}}{2015}]{Winn_2015}
{Winn} J.~N.,  {Fabrycky} D.~C.,  2015, \mn@doi [\araa]
  {10.1146/annurev-astro-082214-122246}, \href
  {https://ui.adsabs.harvard.edu/abs/2015ARA&A..53..409W} {53, 409}

\bibitem[\protect\citeauthoryear{Wu}{Wu}{2019}]{Wu_2019}
Wu Y.,  2019, \mn@doi [\apj] {10.3847/1538-4357/ab06f8}, 874, 91

\bibitem[\protect\citeauthoryear{Yang, Xie  \& Zhou}{Yang
  et~al.}{2020}]{Yang_2020}
Yang J.-Y.,  Xie J.-W.,   Zhou J.-L.,  2020, \mn@doi [\aj]
  {10.3847/1538-3881/ab7373}, 159, 164

\bibitem[\protect\citeauthoryear{{Zhu} \& {Dong}}{{Zhu} \&
  {Dong}}{2021}]{Zhu_2021}
{Zhu} W.,  {Dong} S.,  2021, \mn@doi [\araa]
  {10.1146/annurev-astro-112420-020055}, \href
  {https://ui.adsabs.harvard.edu/abs/2021ARA&A..59..291Z} {59, 291}

\makeatother
\end{thebibliography}

\bsp	
\label{lastpage}

\end{document}